\documentstyle[editedvolume,psfig]{crckapb}

\begin{opening}
\title{INTERPRETING THE MAIN HI AND CO $\ell\!-\!V$ FEATURES
       IN THE GALACTIC BAR FROM SELF-CONSISTENT STELLAR AND
       GAS DYNAMICAL MODELS}

\author{R. FUX}
\institute{Geneva Observatory, CH-1290 Sauverny, Switzerland}

\end{opening}

\runningtitle{THE HI AND CO $\ell\!-\!V$ FEATURES IN THE GALACTIC BAR}

\begin{document}

\vspace{-7.5cm}
\noindent
{\small To appear in: \parbox[t]{6.5cm}{Proc. IAU Symp. 184, {\it The Central
Regions of the Galaxy and Galaxies}, ed. Y. Sofue}}
\vspace{6.7cm}

We propose a new picture accounting for the dominant features in the observed
$\ell-V$ distribution of the Milky Way gas within the central few~kpcs, based
on the gas flow in two self-consistent and symmetry-free barred dynamical
models. These models are snapshots selected from high resolution 3D $N$-body
and SPH simulations with $4\times 10^6$ stellar and dark particles and
$1.5\times 10^5$ gas particles, and with initial conditions adapted from
Fux~(1997). The main results are shown in figure~1. In both models, the
inclination angle of the bar relative to the Sun-Galactic centre line is
$25^{\circ}$, the corotation radius close to 4.4 kpc and the sound speed of
the gas 10~km~s$^{-1}$.
\par Gas dynamics in a rotating bar is driven by shocks which are believed to
generate the dustlanes observed in external barred galaxies (Athanassoula
1992). Such shocks are also expected to exist in the Galaxy and the compressed
gas behind them should produce a typical trace in the HI and~CO $\ell-V$
diagrams. According to our models, the connecting arm is the signature of the
near side branch of these dustlane shocks. The shocked gas plunges onto the
nuclear disc/ring and is responsible for the terminal velocity peak. The far
side branch of the shocks appears as an almost vertical feature in the model
$\ell\!-\!V$ diagram, which is indeed visible in the CO data near
$\ell=-4^{\circ}$. Some other vertical CO features, like the one at
$\ell \approx 5.5^{\circ}$, are understood as molecular clouds crossing the
shock front and therefore undergoing a strong velocity gradient.
\par The 135-km/s arm, long ago suspected as the far side counterpart of the
3-kpc arm, crosses the $\ell=0$ axis at higher absolute velocity than the
latter, because the gas, moving almost parallel to the arm, falls closer to
the nuclear disc/ring and hence reaches larger ``forbidden'' velocities before
crashing into the dustlane shock. This is a typical example of asymmetry
missed by gas flow calculations which do not take into account odd modes.
Note also how well the models are able to reproduce the molecular ring.

\begin{figure}
\centerline{\psfig{figure=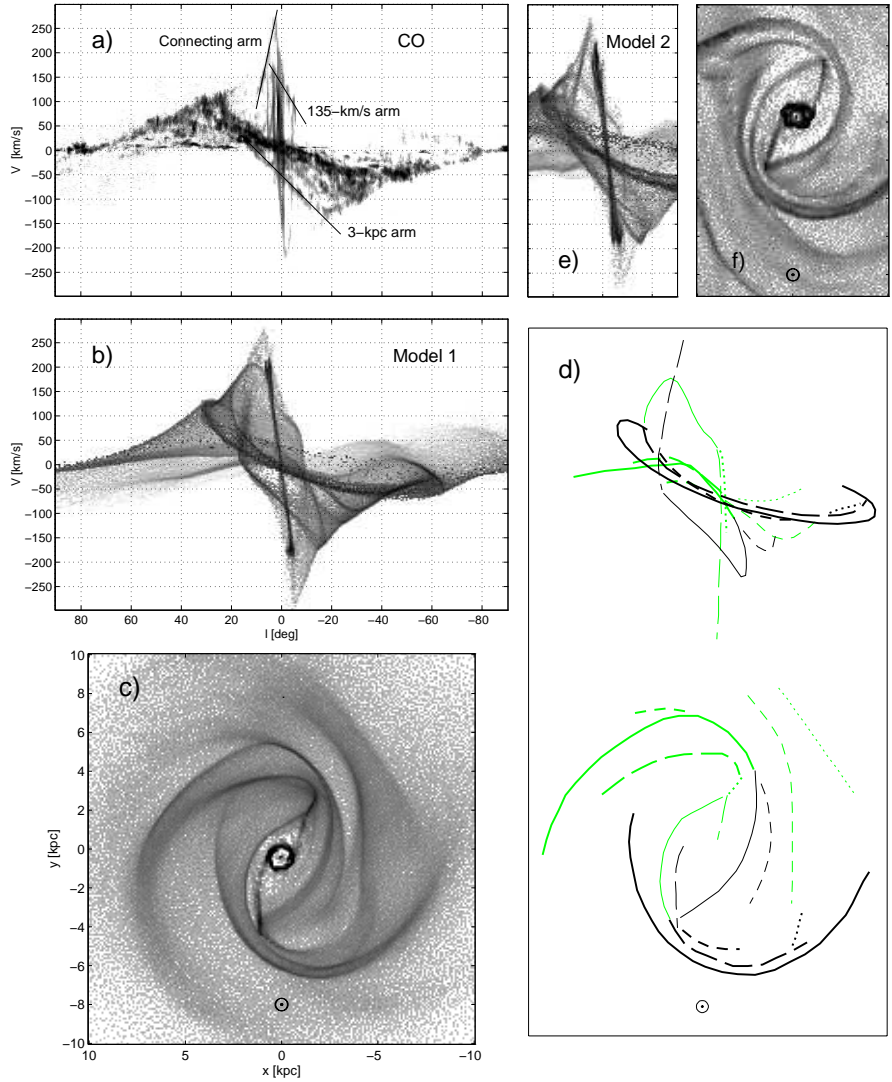,height=14.5cm}}
\caption{a) Observed CO $\ell\!-\!V$ diagram (Dame et al. 1995), with some
relevant features indicated. b-c) Corresponding plot and face-on view for
model 1, which provides a fair global qualitative agreement with the data.
d) Link between the spiral arms and the $\ell\!-\!V$ features in this model.
e-f) Results for model 2, which reproduces almost perfectly the connecting
arm. The location of the Sun is indicated by the $\odot$ symbol.}
\end{figure}


\begin{thebibliography}{}
\bibitem[]{} Athanassoula E. 1992, MNRAS 259, 345
\bibitem[]{} Dame et al. 1995, NCSA Astronomy Digital Image Library
\bibitem[]{} Fux R. 1997, A\&A in press (astro-ph/9706242)
\end{thebibliography}
\end{document}